\documentclass[a4paper,aps,prd,10pt,preprintnumbers,twocolumn,superscriptaddress,nofootinbib,amsmath,amssymb]{revtex4-1}
\usepackage{graphicx}
\usepackage[utf8]{inputenc}
\usepackage[T1]{fontenc}
\usepackage{cmap}
\def\imo{i}

\def\K{{\cal K}}

\begin{document}
\title{Quasinormal modes, stability and shadows of a black hole in the 4D Einstein-Gauss-Bonnet gravity}
\author{R. A. Konoplya} \email{roman.konoplya@gmail.com}
\affiliation{Institute of Physics and Research Centre of Theoretical Physics and Astrophysics, Faculty of Philosophy and Science, Silesian University in Opava, CZ-746 01 Opava, Czech Republic}
\affiliation{Peoples Friendship University of Russia (RUDN University), 6 Miklukho-Maklaya Street, Moscow 117198, Russian Federation}
\author{A. F. Zinhailo} \email{antonina.zinhailo@physics.slu.cz}
\affiliation{Institute of Physics and Research Centre of Theoretical Physics and Astrophysics, Faculty of Philosophy and Science, Silesian University in Opava, CZ-746 01 Opava, Czech Republic}
\begin{abstract}
Recently a $D$-dimensional regularization approach leading to the non-trivial $(3+1)$-dimensional Einstein-Gauss-Bonnet (EGB) effective description  of gravity was formulated which was claimed to bypass the Lovelock's theorem and avoid Ostrogradsky instability. Later it was shown that the regularization is possible only for some broad, but limited, class of metrics and Aoki, Gorji and Mukohyama [arXiv:2005.03859]  formulated a well-defined four-dimensional EGB theory, which breaks the Lorentz invariance in a theoretically consistent and observationally viable way. The black-hole solution of the first naive approach proved out to be also the exact solution of the well-defined theory. Here we calculate quasinormal modes of scalar, electromagnetic and gravitational perturbations and find the radius of shadow for spherically symmetric and asymptotically flat black holes with Gauss-Bonnet corrections. We show that the black hole is gravitationally stable when ($-16 M^2<\alpha \lessapprox 0.6 M^2$). The instability in the outer range is the eikonal one and it develops at high multipole numbers. The radius of the shadow $R_{Sh}$ obeys the linear law with a remarkable accuracy.
\end{abstract}
\maketitle

\section{Introduction}

Quasinormal modes and shadows of black holes are, apparently, among the most interesting characteristics of black holes in the gravitational and electromagnetic spectra.
They have been observed in the modern experiments, still, leaving the wide room for interpretations and alternative theories of gravity \cite{alternative1}.
A number of such alternative theories appeared in attempts to answer a number of fundamental questions which cannot be resolved with General Relativity, such as, for example, construction of quantum gravity or singularity problem. Many of these theories include higher curvature corrections to the Einstein term and one of the most promising approaches is related to the Einstein-Gauss-Bonnet theory (quadratic in curvature) and its Lovelock generalization (for higher than the second order in curvature). In four dimensions, the Einstein-Gauss-Bonnet theory leads to non-trivial corrections of the equations of motion only if the Gauss-Bonnet term is coupled to a matter field, for example, to the dilaton. Various effects in such Einstein-dilaton-Gauss-Bonnet theories were considered in \cite{Blazquez-Salcedo:2016enn,Maselli:2014fca,Ayzenberg:2013wua,Konoplya:2019fpy,Nampalliwar:2018iru,Konoplya:2018arm,Kokkotas:2017ymc,Zinhailo:2019rwd,Cuyubamba:2019qtz,Younsi:2016azx}.
 recently it has been claimed  \cite{Glavan:2019inb} that there is a non-trivial Einstein-Gauss-Bonnet theory of gravity with no extra fields coupled to curvature. There it is stated that there is a general covariant modified theory of gravity in~$D\!=\!4$ space-time dimensions in which  only
the massless graviton propagates  and the theory bypassing the Lovelock's theorem  \cite{Glavan:2019inb} is defined as the limit~$D\!\to\!4$ of the higher dimensional case.
In this singular limit the Gauss-Bonnet invariant produces non-trivial contributions to gravitational dynamics, while preserving the number of graviton degrees of freedom and being free from Ostrogradsky instability. However, in a number of further papers it was shown that the above regularization scheme does not work for an arbitrary metric, so that the above regularization cannot have the status of a well-defined theory \cite{Gurses:2020ofy,Hennigar:2020lsl,Bonifacio:2020vbk,Arrechea:2020evj}.

It was observed the lack of the tensorial description  \cite{Gurses:2020ofy} for the original $4D$ approach \cite{Glavan:2019inb} and found that in some cases different types of regularization lead to the nonuniqueness of some solutions, such as Taub-NUT black holes \cite{Hennigar:2020lsl}. It was pointed out that in four dimensions there is no four-point graviton scattering tree amplitudes other than those leading to the Einstein theory, so that additional degrees of freedom, for instance, a scalar field $(\partial \phi)^4$, should be added for consistency \cite{Bonifacio:2020vbk}. In addition, it was shown that the nonlinear perturbations of the metric cannot be regularized by taking the limit $D \rightarrow 4$ due to divergent terms appearing in the corresponding equations of the Gauss-Bonnet theory \cite{Arrechea:2020evj}. However, such a scalar-tensor approach is based on another scaling limit of the coupling constant and their $D\to 4$ limit is different from the original proposal \cite{Glavan:2019inb}.

In order to solve the above problems another scalar-tensor approach has been proposed by \cite{Lu:2020iav,Kobayashi:2020wqy} and reproduced in \cite{Fernandes:2020nbq,Hennigar:2020fkv} from a different point of view.  This approach implies a four-dimensional Lagrangian and is in the spirit of original proposal of \cite{Glavan:2019inb}:  the works \cite{Lu:2020iav,Kobayashi:2020wqy} first assume a particular ansatz in a higher dimensional spacetime and then take the limit $D \to 4$. However, the problem of this approach is pointed out in \cite{Kobayashi:2020wqy} and discussed in \cite{Bonifacio:2020vbk}: the scalar field is infinitely strongly coupled. These works clarify the essential problem of the original regularization scheme as well as the scalar-tensor approach of \cite{Lu:2020iav,Kobayashi:2020wqy}. The essential problem of the $D\to 4$ limit is the infinitely strong coupling. 

A consistent description of the $4D$ theory has been suggested in \cite{Aoki:2020lig}, where, using the ADM decomposition, the Hamiltonian theory was constructed which breaks the diffeomorphism invariance in an observationally viable way. This well-defined theory does not have an extra scalar degrees of freedom and therefore is free of the problem of infinite coupling. It is essential for our consideration here that the black-hole solution of the original proposal \cite{Glavan:2019inb} \emph{also satisfies} the field equations of the well-defined theory suggested in \cite{Aoki:2020lig}. Moreover, as was shown in a subsequent work by Aoki, Gorji and Mukohyama \cite{Aoki:2020iwm} on the example of the cosmological solution, the dispersion relations for the gravitational perturbations acquire modification in the UV regime. Therefore one could expect that the gravitational spectra in the IR regime, that is, for sufficiently large black holes should be effectively described within the initial simplified proposal of \cite{Glavan:2019inb} once the higher dimensional perturbation equations allow for the dimensional regularization.

An essential requirement for existence of a black hole is its stability against small perturbations of spacetime. The higher dimensional Einstein-Gauss-Bonnet theory is peculiar in this respect: black holes suffer from gravitational instability unless the GB coupling constant is small enough \cite{Dotti:2005sq,Gleiser:2005ra,Konoplya:2017lhs,Konoplya:2017zwo,Takahashi:2010ye,Yoshida:2015vua,Takahashi:2011qda,Gonzalez:2017gwa,Konoplya:2008ix,Takahashi:2012np}. This instability develops at higher multipole numbers and is called, therefore, the eikonal instability  \cite{Dotti:2005sq,Gleiser:2005ra,Konoplya:2017lhs,Konoplya:2017zwo,Takahashi:2010ye,Yoshida:2015vua,Takahashi:2011qda,Gonzalez:2017gwa,Konoplya:2008ix,Cuyubamba:2016cug,Takahashi:2012np}. Usually the eikonal instability essentially constrains the allowed parametric region of black holes. Therefore, it is interesting to know whether such instability exist also for the novel $(3+1)$-dimensional Einstein-Gauss-Bonnet black holes.

Having in mind, first of all, the well-defined 4D Einstein-Gauss-Bonnet theory of Aoki-Gorji-Mukohyama \cite{Aoki:2020lig} we calculate the quasinormal modes of a scalar, electromagnetic and gravitational perturbations with the help of the WKB and time-domain integration methods and find radius of the shadow of an asymptotically flat (3+1)-dimensional Einstein-Gauss-Bonnet black hole. We show that the quasinormal modes are essentially affected by the Gauss-Bonnet coupling. There is the eikonal instability of gravitational perturbations when the coupling constant is not small enough and we find the threshold values of the coupling constant for this instability. We also show that there is no such instability for negative values of the coupling constant. In addition, we discuss the breakdown of the correspondence between the eikonal quasinormal modes and the parameters of the null geodesics formulated in \cite{Cardoso:2008bp}, which is valid here for test fields, but, evidently, not for the gravitational one. We also calculate the radius of a shadow of the 4D Gauss-Bonnet black holes. Our results for the test fields, that is, for test scalar and electromagnetic quasinormal modes as well as for the radius of a shadow must be valid in the full, well-defined theory of Aoki-Gorji-Mukohyama \cite{Aoki:2020lig}, because these effects are essentially propagation of particles and fields in the background black hole metric, and, as such, unlike non-minimally coupled fields, depend only on the form of the black-hole metric, which is the same in all the approaches. The gravitational perturbations studied here, are based on the regularization scheme and, therefore, could only  be considered as an approximation which may be valid in the IR regime, but strongly modified in the UV regime, as it takes place for the cosmological perturbations \cite{Aoki:2020lig}.

The paper is organized as follows. In sec. II we summarize the basic information on the Einstein-Gauss-Bonnet theory and the black hole solution therein. Sec. III is devoted to quasinormal modes of test fields, while Sec. IV discusses the gravitational perturbations, the eikonal instability and the breakdown of the correspondence between the eikonal quasinormal modes and null geodesics. In sec. V we calculate the radius of the shadow of the black hole. Finally, we summarize the obtained results and discuss a number of open questions.

\section{The novel four-dimensional Einstein-Gauss-Bonnet theory and the black hole metric}
 In four dimensional space-time General Relativity is described by the Einstein-Hilbert action,
\begin{equation}
S_{\rm \ss EH}[g_{\mu\nu}]
	= \int \! d^{D\!}x \, \sqrt{-g}
	\left[\frac{M_{\ss \rm P}^2}{2}R\right],
\end{equation}
where $D\!=\!4$ and the reduced Planck mass~$M_{\rm \ss P}$ characterizes the gravitational
coupling strength.
According to the Lovelock's theorem \cite{Lovelock:1971yv,Lovelock:1972vz,
Lanczos:1938sf} General Relativity is the unique four dimensional theory of gravity if one assumes: a) diffeomorphism invariance, b) metricity,  and c) second order equations of motion. In higher than four dimensions the general action satisfying the above conditions is
\begin{equation}
S_{ \rm \ss GB}[g_{\mu\nu}]
	= \int\! d^{D\!}x \, \sqrt{-g} \, \tilde{\alpha} \, \mathcal{G} \, ,
\end{equation}
where~$\alpha$ is a dimensionless (Gauss-Bonnet) coupling constant and~$\mathcal{G}$ is the
Gauss-Bonnet invariant,~$\mathcal{G} \!=\!
	{R^{\mu\nu}}_{\rho\sigma} {R^{\rho\sigma}}_{\mu\nu}
	\!-\! 4 {R^\mu}_\nu {R^\nu}_\mu \!+\! R^2 \!=\!
	6 {R^{\mu\nu}}_{[\mu\nu} {R^{\rho\sigma}}_{\rho\sigma]}$.
The idea suggested in \cite{Glavan:2019inb} is to rescale the coupling constant,
\begin{equation}
\alpha \to \tilde{\alpha}/(D\!-\!4) \, ,
\label{coupling}
\end{equation}
of the Gauss-Bonnet term, and only afterwards to consider the limit~$D\!\rightarrow\!4$. This leads to the solution for a static and spherically
symmetric case in an arbitrary number of dimensions
dimensions~$D\!\ge\!5$,
\begin{equation}\label{sphansatz}
ds^2 = -f(r)dt^2
	+f^{-1}(r)dr^2
	+r^2d\Omega_{D-2}^2
\end{equation}
which was  already found in Ref.~\cite{Boulware:1985wk} (see also \cite{Wheeler,Wiltshire:1985us,Cai:2001dz}). This solution is extended to~$D\!=\!4$ solutions via the re-scaling prescribed in \cite{Glavan:2019inb}, and then by taking the limit~$D\!\rightarrow\!4$,

\begin{equation}
\label{sch-de}
f(r)	= 1 + \frac{r^2}{32 \pi \tilde{\alpha} G}
	\Biggl[ 1\pm \biggr( \!1 \!+\! \frac{128 \pi\tilde{\alpha} G^2 M}{r^3} \biggr)^{\!\! 1/2 \,} \!
	\Biggr].
\end{equation}
Here the Newton's constant is ~$G \!=\! 1/(8\pi M_{\ss \rm P}^2)$ and $M$ is a mass parameter. A similar effective approach for constructing the black-hole metric in the charged case and for higher order Lovelock corrections have been recently suggested in \cite{Fernandes:2020rpa,Konoplya:2020qqh}.

An essential moment for our future consideration is that this solution is not only the result of the dimensional regularization suggested in \cite{Glavan:2019inb}, but also an exact solution of the well-defined truly four-dimensional Aoki-Gorji-Mukohyama theory \cite{Aoki:2020lig}, which
allows  for Hamiltonian description and uses the ADM decomposition \cite{Aoki:2020lig}. This theory breaks the Lorenz invariance via modification of the dispersion relations in the UV regime, making the whole approach consistent with the current observations in the IR regime.

Therefore, all the effects related to not strongly coupled fields, but to fields propagating in the black hole background can be safely studied independently of the incompleteness of the theory suggested in \cite{Glavan:2019inb}, because the same black-hole solution is valid also in the above well-defined theory \cite{Aoki:2020lig}. In this approach there is no scalar propagating degree of freedom \cite{Lu:2020mjp} and the dispersion relations obtain corrections in the UV part of the spectrum, leaving the astrophysically relevant IR limit of the spectrum unaffected \cite{Aoki:2020iwm}. Therefore, the gravitational perturbations treated via the same dimensional regularization as in \cite{Glavan:2019inb}, are expected  be valid within the full theory \cite{Aoki:2020iwm}  for analysis of the IR part of the spectrum, that is, when the black hole is of order of the radiation's wavelength or larger.

The solution of the field equations has two branches corresponding to different signs in front of the square root.
Here we will study ``the minus'' case of the above metric, as it leads to the asymptotically flat solution, unlike ``the plus'' case, which is effectively asymptotically de Sitter one in the absence of the cosmological constant.
If $\tilde{\alpha}\!<\!0$, there is no real solution for $r^3\!<\!-128\pi \tilde{\alpha} G^2 M$, which means invalidity of the solution in the central region at some distance from the center  which, anyway, is hidden under the event horizon \cite{Guo:2020zmf} for sufficiently small absolute values of the coupling constant. Mostly, we will consider the $\tilde{\alpha} >0$ case here. Nevertheless,  as the metric for negative $\tilde{\alpha}$ is well-behaved outside the event horizon, we will consider the form of the effective potentials, stability regions and obtain some results on quasinormal modes, which are valid for negative $\alpha$ as well.

The event horizon is the larger root of the following ones:
\begin{equation}
r^{\ss H}_\pm = GM \Biggl[ 1 \pm\sqrt{1 - \frac{16 \pi \tilde{\alpha}}{GM^2}} \ \Biggr] \, .
\end{equation}
For negative values of $\tilde{\alpha}$ there is only one horizon, corresponding to the ``the plus'' sign in the above relation.
Notice also that the above black-hole metric was considered earlier in the context of corrections to the entropy formula in \cite{Cognola,Cai:2009ua}. From here and on we will consider $32 \pi \tilde{\alpha}$ as a new coupling constant $\alpha$
\begin{equation}
\alpha = 32 \pi \tilde{\alpha},
\end{equation}
and use the units $G=1$ and $M=1/2$.
Now we are in position to consider quasinormal modes and shadows of the above black holes.

\begin{figure*}
\resizebox{\linewidth}{!}{\includegraphics*{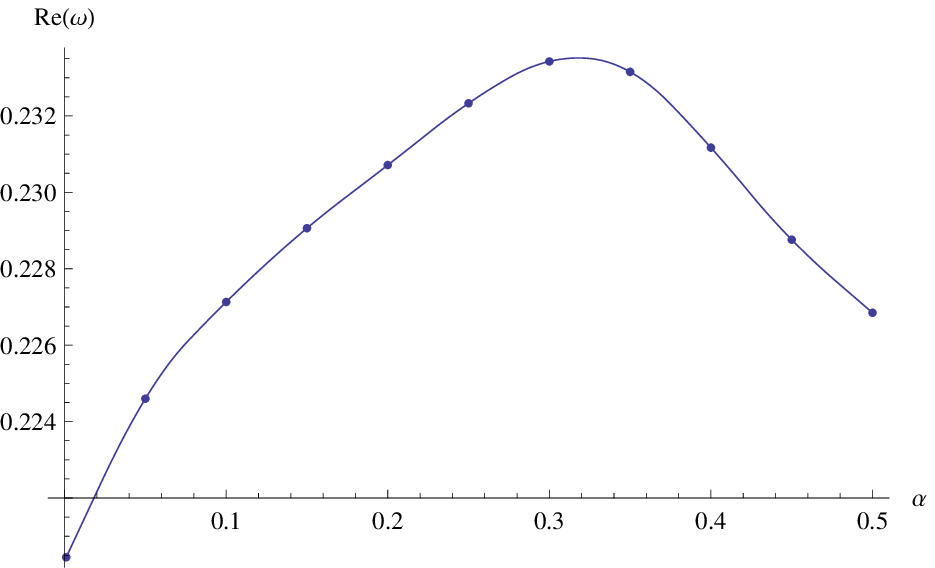}\includegraphics*{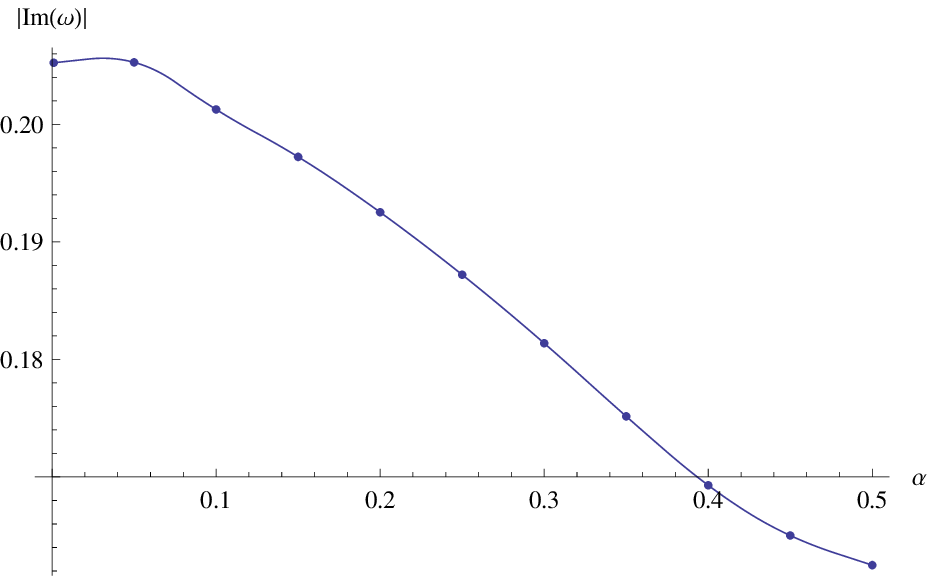}}
\caption{The fundamental ($n=0$) quasinormal mode computed by the WKB approach for $\ell=0$ scalar perturbations as a function of $\alpha$, $M=1/2$.}\label{fig1}
\end{figure*}

\begin{figure*}
\resizebox{\linewidth}{!}{\includegraphics*{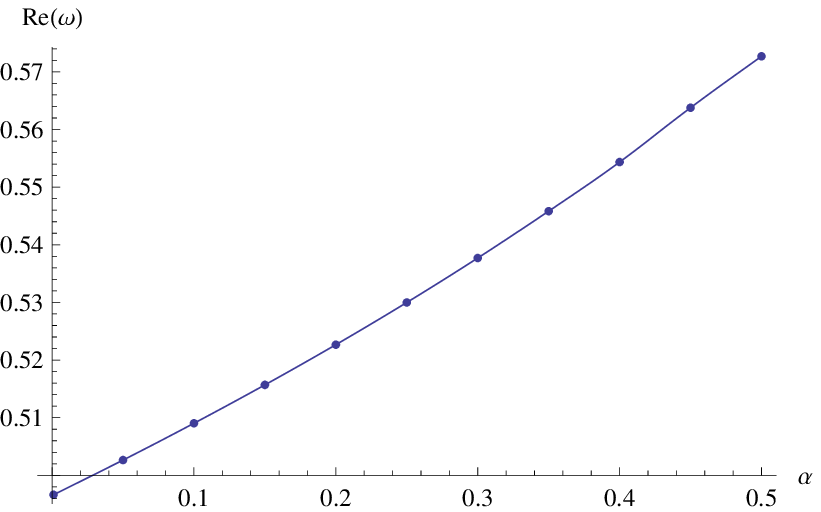}\includegraphics*{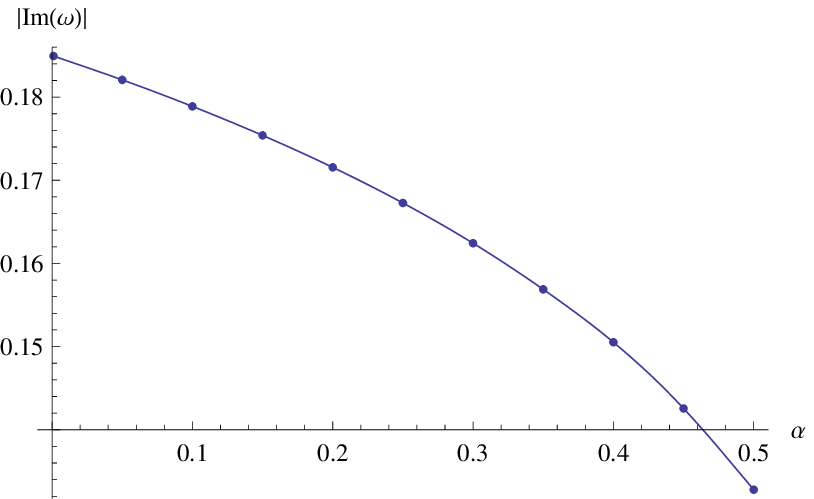}}
\caption{The fundamental ($n=0$) quasinormal mode computed by the WKB approach for $\ell=1$ electromagnetic perturbations as a function of $\alpha$, $M=1/2$.}\label{fig2}
\end{figure*}

\section{Quasinormal modes of scalar and electromagnetic fields}

Taking into account the re-definition of the coupling constant, the metric function has the form
\begin{equation}
\label{sch-de2}
f(r)	= 1 + \frac{r^2}{\alpha }
	\Biggl[ 1 - \biggr( \!1 \!+\! \frac{4 \alpha M}{r^3} \biggr)^{\!\! 1/2 \,} \!
	\Biggr].
\end{equation}
The exterior black hole solution exists for values of the coupling constant $\alpha$ being in the range
\begin{equation}
- 16 M^2 <\alpha< 2 M^2.
\end{equation}
Let us notice, that our $\alpha$ differs from those of \cite{Guo:2020zmf,Konoplya:2020juj} by a factor $2$.
The general covariant equation for a massless scalar field has the form
\begin{equation}\label{KGg}
\frac{1}{\sqrt{-g}}\partial_\mu \left(\sqrt{-g}g^{\mu \nu}\partial_\nu\Phi\right)=0,
\end{equation}
and for an electromagnetic field it has the form
\begin{equation}\label{EmagEq}
\frac{1}{\sqrt{-g}}\partial_\mu \left(F_{\rho\sigma}g^{\rho \nu}g^{\sigma \mu}\sqrt{-g}\right)=0\,,
\end{equation}
where $\mu, \nu =0, 1, 2, 3$ and $F_{\rho\sigma}=\partial_\rho A_{\sigma}-\partial_\sigma A_{\rho}$ and $A_\mu$ is a vector potential.

Since the background is spherically symmetric,
we can expand $A_{\mu}$ in terms of the vector spherical
harmonics (see \cite{Ruffini}):
\begin{widetext}
{\small
\begin{eqnarray}
A_{\mu}(t,r,\theta,\phi)=\sum_{\ell , m}\left( \begin{array}{cc}\left[
 \begin{array}{c} 0 \\ 0 \\
 \frac{a^{\ell m}(t,r)}{\sin\theta}\partial_\phi Y_{\ell m}\\
 -a^{\ell m}(t,r)\sin\theta\partial_\theta Y_{\ell m}\end{array}\right] &
 +\left[ \begin{array}{c}f^{\ell m}(t,r)Y_{\ell m}\\h^{\ell m}(t,r)Y_{\ell m} \\
 k^{\ell m}(t,r) \partial_\theta Y_{\ell m}\\ k^{\ell m}(t,r) \partial_\phi
 Y_{\ell m}\end{array}\right] \end{array}\right)\,,
\label{expansion}
\end{eqnarray}}
\end{widetext}
\noindent where the first term in the right-hand side has parity $(-1)^{\ell+1}$
and the second term has parity $(-1)^{\ell}$, $m$ is the azimuthal number
and $\ell$ is the angular quantum number.  If we put this expansion into Maxwell's
equations (\ref{EmagEq}) we get a second order differential
equation for the perturbation:
\begin{equation}
\frac{\partial^{2} \Psi(r)}{\partial r_*^{2}} +
\left\lbrack\omega^2-V(r)\right\rbrack\Psi(r)=0 \,,
\label{wavemaxwell}
\end{equation}
where the wavefunction $\Psi(r)$ is a linear combination of the functions
$f^{\ell m}$, $h^{\ell m}$, $k^{\ell m}$ and $a^{\ell m}$ as appearing in
(\ref{expansion}). The form of $\Psi$
depends on the parity: for odd parity, i.e, $(-1)^{\ell+1}$, $\Psi$
is explicitly given by $\Psi=a^{\ell m}$ whereas for even parity $(-1)^{\ell}$
it is given by $\Psi=\frac{r^2}{\ell(\ell+1)}\left(-i\omega
h^{\ell m}-\frac{df^{\ell m}}{dr}\right)$.
It is assumed that the time dependence is $\Psi(t,r)=e^{-i\omega t}\Psi(r)$.

The case of a scalar field is simpler and implies expansion in terms of the standard spherical harmonics. Summarizing, after separation of the variables equations (\ref{KGg},\ref{EmagEq}) take the following form
\begin{equation}\label{wave-equation}
\frac{d^2\Psi_s}{dr_*^2}+\left(\omega^2-V_{s}(r)\right)\Psi_s=0,
\end{equation}
where $s=scal$ corresponds to a scalar field and $s=em$ to the electromagnetic field. The ``tortoise coordinate'' $r_*$ is defined by the relation
$ dr_*=dr/f(r)$, and the effective potentials are
\begin{equation}\label{scalarpotential}
V_{scal}(r) = f(r)\left(\frac{\ell(\ell+1)}{r^2}+\frac{1}{r}\frac{d f(r)}{dr}\right),
\end{equation}
\begin{equation}\label{empotential}
V_{em}(r) = f(r)\frac{\ell(\ell+1)}{r^2}.
\end{equation}
The effective potentials have the form of a positive definite potential barrier with a single maximum.
Quasinormal modes $\omega_{n}$ correspond to solutions of the master wave equation (\ref{wave-equation}) with the requirement of the purely outgoing waves at infinity and purely incoming waves at the event horizon (see,  for example, \cite{Konoplya:2011qq,Kokkotas:1999bd}):
\begin{equation}
\Psi_{s} \sim \pm e^{\pm i \omega r^{*}}, \quad r^{*} \rightarrow \pm \infty.
\end{equation}

In order to find quasinormal modes we shall use the two independent methods:
\begin{enumerate}
\item Integration of the wave equation (before introduction the stationary ansatz) in time domain at a given point in space \cite{Gundlach:1993tp}.
We shall integrate the wave-like equation rewritten in terms of the light-cone variables $u=t-r_*$ and $v=t+r_*$. The appropriate discretization scheme was suggested in \cite{Gundlach:1993tp}:
$$
\Psi\left(N\right)=\Psi\left(W\right)+\Psi\left(E\right)-\Psi\left(S\right)-
$$
\begin{equation}\label{Discretization}
-\Delta^2\frac{V\left(W\right)\Psi\left(W\right)+V\left(E\right)\Psi\left(E\right)}{8}+{\cal O}\left(\Delta^4\right)\,,
\end{equation}
where we used the following notation for the points:
$N=\left(u+\Delta,v+\Delta\right)$, $W=\left(u+\Delta,v\right)$, $E=\left(u,v+\Delta\right)$ and $S=\left(u,v\right)$. The initial data are given on the null surfaces $u=u_0$ and $v=v_0$.

\item In the frequency domain  we will use the WKB method of Will and Schutz \cite{Schutz:1985zz}, which was extended to higher orders in \cite{Iyer:1986np,Konoplya:2003ii,Matyjasek:2017psv} and made even more accurate by the usage of the Pade approximants in \cite{Matyjasek:2017psv,Hatsuda:2019eoj}.
The higher-order WKB formula \cite{Konoplya:2019hlu}:
$$ \omega^2=V_0+A_2(\K^2)+A_4(\K^2)+A_6(\K^2)+\ldots- $$
\begin{equation}\nonumber
\imo \K\sqrt{-2V_2}\left(1+A_3(\K^2)+A_5(\K^2)+A_7(\K^2)\ldots\right),
\end{equation}
where $\K$ takes half-integer values. The corrections $A_k(\K^2)$ of order $k$ to the eikonal formula are polynomials of $\K^2$ with rational coefficients and depend on the values of higher derivatives of the potential $V(r)$ in its maximum. In order to increase accuracy of the WKB formula, we follow Matyjasek and Opala \cite{Matyjasek:2017psv} and use Padé approximants.
\end{enumerate}

As both methods are very well known (\cite{Konoplya:2019hlu,Konoplya:2011qq}), we will not describe them in this paper in detail, but will simply show that both methods are in a good agreement in the common range of applicability.

\begin{table}
\begin{tabular}{p{1cm}cccc}
\hline
$\alpha$ & WKB (6th order, $\tilde{m} =5$) & Time-domain  \\
\hline
0.001    & $0.585950-0.195270 i$ &  $0.587327 - 0.195509 i$  \\
0.05    & $0.590797-0.192066 i$ & $0.592067 - 0.192181 i$  \\
0.1    & $0.595897-0.188579 i$ &  $0.597066 - 0.188587 i$  \\
0.15    & $0.601189-0.184841 i$ & $0.602249 - 0.184756 i$  \\
0.2    & $0.606692-0.180804 i$ &  $0.607636 - 0.180642 i$  \\
0.25    & $0.612426-0.176402 i$ & $0.613249 - 0.176179 i$  \\
0.3    & $0.618405-0.171552 i$  & $0.619107 - 0.171282 i$  \\
0.35    & $0.624629-0.166145 i$ & $0.625217 - 0.165835 i$ \\
0.4    & $0.631108-0.160083 i$ &  $0.631559 - 0.159674 i$  \\
0.45    & $0.637965-0.152884 i$ & $0.638048 - 0.152586 i$  \\
0.5    & $0.644336-0.144444 i$ &  $0.644465 - 0.144330 i$  \\
\hline
\end{tabular}
\caption{The fundamental quasinormal mode of the scalar field ($\ell=1$, $n=0$, $M =1/2$) as a function of $\alpha$. }\label{tab4}
\end{table}

\begin{figure}
\centerline{\resizebox{\linewidth}{!}{\includegraphics*{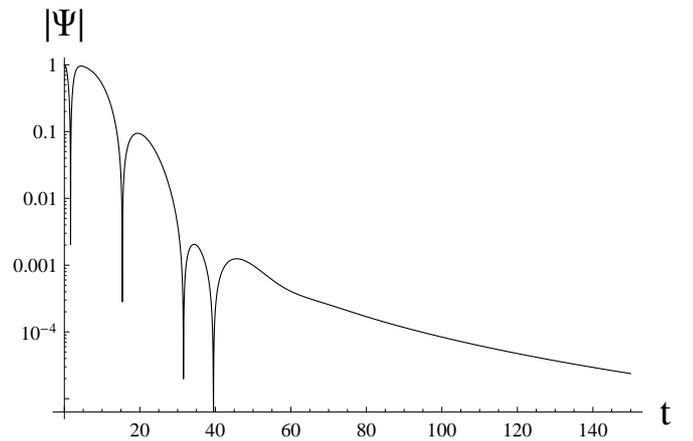}}}
\caption{The time-domain profile of the $\ell=0$ scalar perturbations, $\alpha = 0.1$, $M =1/2$.}\label{fig3}
\end{figure}

\begin{figure}
\centerline{\resizebox{\linewidth}{!}{\includegraphics*{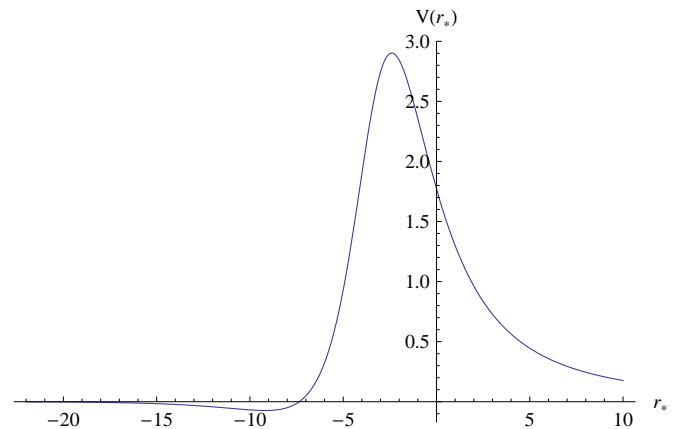}}}
\caption{The effective potnetial for vector type of gravitational perturbations $\ell=5$, $\alpha = 0.45$, $M =1/2$.}\label{fig7}
\end{figure}

\begin{figure}
\centerline{\resizebox{\linewidth}{!}{\includegraphics*{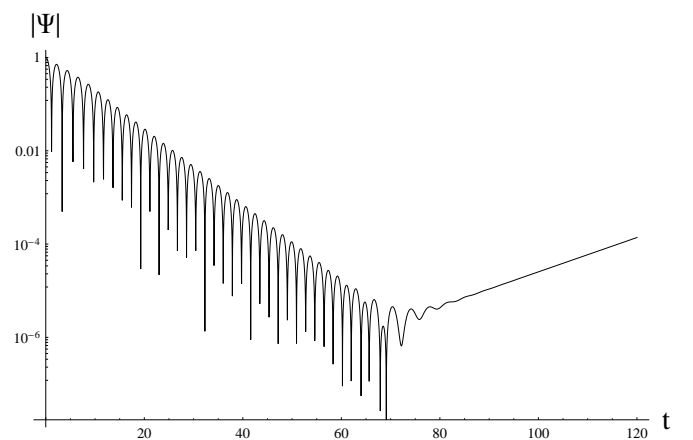}}}
\caption{Time-domain profile for vector type of gravitational perturbations $\ell=5$, $\alpha = 0.45$, $M =1/2$.}\label{fig8}
\end{figure}

From the table I and figure \ref{fig2} one can see that when increasing the coupling constant $\alpha$, the real oscillation frequency of $\ell>0$ modes  is monotonically increased, while the damping rate is decreased. The behavior of the lowest scalar multipole $\ell=0$
is different according to the WKB data given on fig. \ref{fig1} and one could suspect that there is lacking accuracy of the WKB technique. However, the time-domain calculations show qualitatively similar behavior: the real oscillation frequency begins to decrease at some value of $\alpha$. In the next section we will show that at these values of $\alpha$ when the $Re (\omega)$ is non-monotonic, the gravitational instability develops, so that no real black hole can exist.
At the same time, even the time-domain data cannot be fully trusted for $\ell=0$, as the extraction of the frequency is difficult in this case, because the quasinormal ringing occurs only during a few oscillations and then goes over into the asymptotic tails (see fig. \ref{fig3}). When using the WKB method we applied the Pade approximants as prescribed in \cite{Matyjasek:2017psv} and used the 6th WKB order with $\tilde{m} =5$ \cite{Konoplya:2019hlu}.
Unlike the lowest $\ell=0$ case, quasirnomal modes for higher multipoles calculated by the WKB and time-domain integration methods are in a very good agreement, what will be illustrated for gravitational perturbations in the next section.

\begin{figure}
\centerline{\resizebox{\linewidth}{!}{\includegraphics*{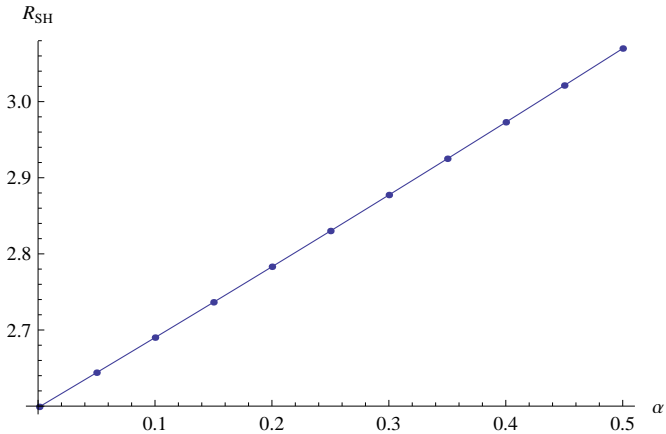}}}
\caption{Radius of the shadow as a function of $\alpha$, $r_{+}=1$.}\label{fig4}
\end{figure}

In the regime of high multipole numbers the WKB formula of the first order can be applied
$$
\omega^2=V_0+\sqrt{-2V^{\prime \prime}_{0}}\left(n+\frac{1}{2}\right)i,
$$
where $V_{0}$ is the peak of the effective potential and $V^{\prime \prime}_{0}$ its second derivative at the peak. This formula can be expanded in terms of $1/\ell$ (see \cite{Churilova:2019jqx} for a general approach).
The peak of the effective potential has the following form:
\begin{equation}
r_{0} = 3 M-\frac{2 \alpha}{9 M} + {\cal O}\left({\alpha^2}\right),
\end{equation}
while the quasinormal frequencies in this regime are

$$ \omega =  \frac{1}{3 \sqrt{3} M} \left( \ell +\frac{1}{2} - i \left(n+\frac{1}{2}\right) \right)+ $$
\begin{equation}
\frac{\alpha}{81 \sqrt{3} M^3} \left(\ell + \frac{1}{2} + i (2 n+1) \right)+O\left(1/\ell, \alpha ^2\right)
\end{equation}
Let us notice that the above eikonal formula is valid for both positive and negative $\alpha$, whenever the black-hole solution under consideration is stable against gravitational perturbations. When the coupling $\alpha$ vanishes, the above formula goes over into the well-known expression for the Schwarzschild limit \cite{MashhoonBlome}.

\section{Gravitational perturbations and the eikonal instability}

\subsection{The perturbation equations}
In \cite{Takahashi:2010ye} it was shown that after the decoupling of angular variables and some algebra, the gravitational perturbation equations of the higher dimensional Einstein-Gauss-Bonnet theory can be reduced to the second-order master differential equations
\begin{equation}\label{21}
\left(\frac{\partial^2}{\partial t^2}-\frac{\partial^2}{\partial r_*^2}+V_i(r_*)\right)\Psi_{i}(t,r_*)=0,
\end{equation}
where $\Psi_i$ are the wave functions, $r_*$ is the tortoise coordinate,
\begin{equation}
dr_*\equiv \frac{dr}{f(r)}=\frac{dr}{1-r^2\psi(r)},
\end{equation}
and $i$ stands for $v$ (\emph{vector}), and $s$ (\emph{scalar}) types of gravitational perturbations. These perturbations transforms as scalars and vector respectively the rotation group on a $(D-2)$-sphere. They should not be confused with the test scalar or vector   considered in the previous section. The vector type of gravitational pertubations is also called the axial type, and the scalar is known as the polar type. From the analysis of gravitational perturbations of the tensor type of the cosmological solution in the full theory \cite{Aoki:2020iwm} we see that only the tensor type of perturbations acquires corrections in comparison with the result obtained via the naive regularization scheme  \cite{Glavan:2019inb}. This correction is expressed in eq. (33) of \cite{Aoki:2020iwm} and the correction is of the order $\alpha H^2$, where $H$ is inverse proportional to the characteristic length scale. For the black hole case this means that the correction which we possible neglect must of the order of $\alpha/M^2$, where $M$ is the black hole mass. Therefore, the equation (\ref{21}) should be a good approximation to the one for the full theory, when the Gauss-Bonnet coupling is much smaller than the typical length scale, namely $|\alpha| \ll M^2$.  Nevertheless, there remains a chance that, as in the case of he cosmological background, no new corrections will appear due the full theory, because perturbation equations for the scalar and tensor modes of the cosmological background in the well defined theory \cite{Aoki:2020iwm}  are the same as in the naive regularization scheme \cite{Glavan:2019inb}.

The explicit forms of the effective potentials $V_s(r)$, $V_v(r)$ are given by
\begin{eqnarray}\nonumber
V_v(r)&=&\frac{(\ell-1)(\ell+n)f(r)T'(r)}{(n-1)rT(r)}+R(r)\frac{d^2}{dr_*^2}\Biggr(\frac{1}{R(r)}\Biggr),\\\nonumber
V_s(r)&=&\frac{2\ell(\ell+n-1)f(r)P'(r)}{nrP(r)}+\frac{P(r)}{r}\frac{d^2}{dr_*^2}\left(\frac{r}{P(r)}\right),
\end{eqnarray}
where $n=D-2$, $\ell=2,3,4,\ldots$ is the multipole number, $T(r)$ is given in \cite{Takahashi:2010ye}, and
$$R(r)=r\sqrt{|T'(r)|},$$ $$P(r)=\frac{2(\ell-1)(\ell+n)-nr^3\psi'(r)}{\sqrt{|T'(r)|}}T(r).$$
For large $\ell$ the effective potentials can be approximated as follows:
\begin{equation}
V_{i} = \ell^2 \left(\frac{f_{i}(r)}{r^2} + {\cal O}\left(\frac{1}{\ell}\right)\right),
\end{equation}
where, $i$ stands for vector (v)  and scalar (s) types of gravitational perturbations. Here, we have
\begin{equation}\label{fv}
f_{v}(r) = \frac{f(r)r T'(r)}{(D-3)T(r)},
\end{equation}
\begin{equation}\label{fs}
f_{s}(r) = \frac{rf(r)(2T'(r)^2-T(r)T''(r))}{(D-2) T'(r)T(r)}.
\end{equation}

One can see that the higher dimensional field equations after the re-scaling $\alpha \rightarrow \alpha/(D-4)$ do not contain a singular factor in the limit $D \rightarrow 4$, so that not only for the background black hole metric, but also the perturbation equations for the time-dependent metric can be regularized in the same way as in  \cite{Glavan:2019inb}. In other words, we can use the master equation obtained for the higher dimensional Einstein-Gauss-Bonnet case implying an arbitrary background metric function $f(r)$ and then perform the re-scaling $\alpha \rightarrow \alpha/(D-4)$  in it.

The effective potentials have the form of the potential barrier in this case, but, as in the higher dimensional EGB gravity, with a negative gap near the event horizon at larger values of the coupling constant $\alpha$. This negative gap becomes infinite when the multipole number $\ell$ goes to infinity, which means the so called eikonal instability \cite{Dotti:2005sq,Gleiser:2005ra,Konoplya:2017lhs,Konoplya:2017zwo,Takahashi:2010ye,Yoshida:2015vua,Takahashi:2011qda,Gonzalez:2017gwa,Konoplya:2008ix,Takahashi:2012np}.
Indeed, the effective potential for the vector type of gravitational perturbations has the following form in the eikonal regime:
\begin{widetext}
\begin{equation}
V_v = \frac{32 K \ell ^2 \left(r^3-2 M \alpha \right) \left(4 M \alpha +r^3\right) \left(r
   \sqrt{\frac{4 M \alpha }{r^3}+1}-4 M+r\right)}{r^3 \left(\sqrt{\frac{4 M \alpha
   }{r^3}+1}+1\right) \left(r^3 \left(\sqrt{\frac{4 M \alpha }{r^3}+1}+1\right)+2 M \alpha
   \right)^2 \left(r^3 \left(\sqrt{\frac{4 M \alpha }{r^3}+1}+1\right)+4 M \alpha \right)^4} +{\cal O}\left(\frac{1}{\ell}\right),
\end{equation}
\end{widetext}
where
$$K=2 M^4 \alpha ^4+4 M^3 r^3 \alpha ^3 \left(\sqrt{\frac{4 M \alpha }{r^3}+1}+4\right)+$$
$$ 10 M^2 r^6  \alpha ^2 \left(\sqrt{\frac{4 M \alpha }{r^3}+1}+2\right)+ r^{12} \left(\sqrt{\frac{4 M \alpha
   }{r^3}+1}+1\right)+$$
\begin{equation}
M r^9 \alpha  \left(6 \sqrt{\frac{4 M \alpha }{r^3}+1}+8\right).
\end{equation}

The effective potential for the scalar type of gravitational perturbations (see fig. \ref{fig9}) is too cumbersome and, therefore, we do not write it down here explicitly.

\subsection{The (in)stability region}

According to the analysis of the detailed eikonal instability given in \cite{Gleiser:2005ra,Takahashi:2011qda,Takahashi:2010ye,Konoplya:2017lhs} for the higher-dimensional Einstein-Gauss-Bonnet theory, once for some fixed $\alpha$ the negative gap appears and becomes deeper at higher multipoles, there is some sufficiently large $\ell$ for which the bound state with negative energy appears, which means the onset of instability. Therefore,  investigation of regions in which the effective potential is positive definite not only at the lower, but also at high multipoles is sufficient to determine the stability, while the negative gap becoming deeper when $\ell$ is increased signifies the eikonal instability.

Thus, by looking at the parametric regions in which the effective potentials are positive or negative, one can see that the re-scaled potential has the following region of the eikonal stability:
\begin{equation}
\alpha \lessapprox 1.57 M^2, \quad (vector~ type)
\end{equation}
while for smaller $\alpha$ the black hole is stable against vector perturbations. This type of instability was called the ghost instability in  \cite{Takahashi:2010gz}.
The scalar type of gravitational perturbations imposes even stronger bound on the coupling constant:
\begin{equation}
\alpha \lessapprox 0.6 M^2, \quad (scalar~ type)
\end{equation}

The profile of the quasinormal ringing for gravitational perturbations representing a typical time-domain evolution of instability is shown on fig. \ref{fig8}: as for the higher dimenaional EGB theory \cite{Konoplya:2008ix} it develops after a long period of damped quasinormal oscillations for every finite $\ell$ and the eikonal regime $\ell \rightarrow \infty$ corresponds to the parametrically largest region of instability.

\begin{table}
\begin{tabular}{|c|c|c|}
  \hline
  $\alpha$ & $QNM$ (WKB) & $QNM$ (Time-domain) \\
  \hline
  $-1.9$ &$1.144885 - 0.528724  i$ & $1.14419 - 0.549234 i$ \\
  $-1.5$ &$0.977526 - 0.418250 i$ & $0.947443 - 0.420511 i$ \\
  $-1.0$ &$0.888974 - 0.270908 i$ & $0.875314 - 0.269414 i$ \\
  $-0.9$ &$0.865131 - 0.261466 i$ & $0.860759 - 0.250131 i$ \\
  $-0.8$ &$0.840979 - 0.247055 i$ & $0.846653 - 0.233332 i$ \\
  $-0.7$ &$0.820798 - 0.230881 i$ & $0.832837 - 0.218772 i$ \\
  $-0.6$ &$0.813614 - 0.212003 i$ & $0.819222 - 0.206298 i$ \\
  $-0.5$ &$0.805949 - 0.196027 i$ & $0.805771 - 0.195845 i$ \\
  $-0.4$ &$0.792742 - 0.188256 i$ & $0.792508 - 0.187443 i$ \\
  $-0.3$ &$0.780243 - 0.181006 i$ & $0.779553 - 0.181222 i$ \\
  $-0.2$ &$0.767666 - 0.177550 i$ & $0.767195-  0.177412 i$ \\
  $-0.1$ &$0.756476 - 0.176684 i$ & $0.755645 - 0.176363 i$ \\
  $-0.001$ &$0.747334 - 0.177826 i$ & $0.747018 - 0.177986 i$ \\
  $0.001$ & $0.747146 - 0.177894 i$ & $0.74727 - 0.177938 i$ \\
  $0.05$ & $0.744343 - 0.179866 i$ & $ 0.744444 - 0.179523 i$ \\
  $0.10$ & $0.743063 - 0.181674 i$ & $0.742834 - 0.181299 i$\\
  $0.15$ & $0.743022 - 0.182724 i$ & $0.742599 - 0.182762 i$\\
  \hline
\end{tabular}
\caption{Gravitational quasinormal modes the vector (axial) type for various values of the coupling constant $\alpha$ in the stability sector; $\ell=2$. At $\alpha \approx -1.5$ and smaller there are two concurrent modes in the spectrum and the other mode (not shown in the table) is $\omega = 0.676472 - 0.506277 i$. The corresponding time-domain profile is shown on fig. \ref{fig10}.}
\end{table}
\begin{figure}
\centerline{\resizebox{\linewidth}{!}{\includegraphics*{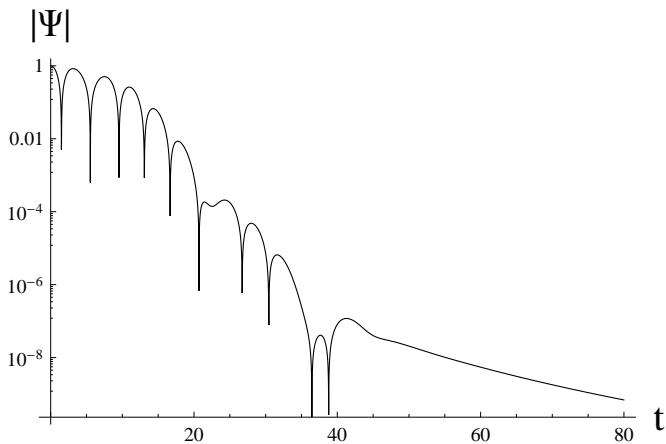}}}
\caption{Time-domain profile for the vector (axial) type of gravitational perturbations $\ell=2$ for $\alpha = -1.9$. The two dominating modes are $\omega_{0} =0.676472 - 0.506277 i$ and $\omega_{1} =1.14419 - 0.549234 i$, $M=1/2$.}\label{fig10}
\end{figure}

\begin{table}
\begin{tabular}{|c|c|c|}
  \hline
  $\alpha$ & $QNM$ (WKB) & $QNM$ (Time-domain) \\
  \hline
  $-1.9$ &$ 1.65941 - 0.51949 i$ & $1.66368 - 0.522244 i$ \\
  $-1.5$ &$1.17625 -  0.31545 i$ & $1.47459 - 0.431442 i$ \\
  $-1.0$ &$.261665 - 0.345704  i$ & $1.26762 - 0.338142 i$ \\
  $-0.9$ &$1.219245 - 0.330473 i$ & $1.22271 - 0.334084 i $ \\
  $-0.8$ &$1.176251 - 0.315445 i$ & $1.17344 - 0.317964 i$ \\
  $-0.7$ &$1.132342 - 0.300441 i$ & $1.12973 - 0.299166 i$ \\
  $-0.6$ &$1.087276 - 0.285259 i$ & $1.0864 - 0.283879 i$ \\
  $-0.5$ &$1.041801 - 0.267132 i$ & $1.03887 - 0.269181 i$ \\
  $-0.4$ &$0.988244 - 0.251233 i$ & $0.988105 - 0.252275 i$ \\
  $-0.3$ &$0.934963 - 0.233770 i$ & $0.93482 - 0.233821 i$ \\
  $-0.2$ &$0.877534 - 0.214423 i$ & $0.877592 - 0.214429  i$ \\
  $-0.1$ &$0.815283 - 0.194448  i$ &  $0.815284 - 0.194505 i$ \\
  $-0.001$ &$0.748051 - 0.178045 i$ & $0.747528 - 0.178118 i$ \\
  $0.001$ & $0.746635 - 0.177809 i$ & $ 0.746114 - 0.177876 i$ \\
  $0.05$ & $0.711526 - 0.174330 i$ & $ 0.711232 - 0.174252 i$ \\
  $0.10$ & $0.677977 - 0.177305 i$ & $ 0.677517 - 0.176864 i$\\
  $0.15$ & $0.651891 - 0.184149 i$ & $0.651344 - 0.184509 i$\\
  \hline
\end{tabular}
\caption{Gravitational quasinormal modes of the scalar (polar) type for various values of the coupling constant $\alpha$ in the stability sector; $\ell=2$, $M=1/2$. At $\alpha \approx -1.5$ and smaller the two concurrent modes appear, which makes the agreement between WKB and time-domain integration worse.}
\end{table}

\begin{figure}
\centerline{\resizebox{\linewidth}{!}{\includegraphics*{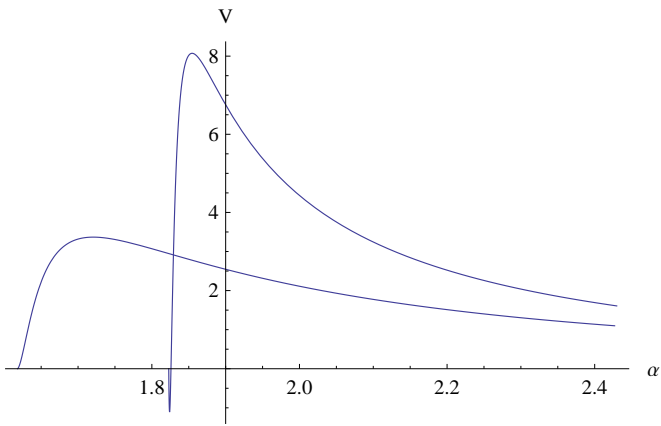}}}
\caption{Effective potentials for the scalar type of gravitational perturbations $\ell=2$ for $\alpha = -3$ (with a negative gap) and $\alpha =-2$ (positive definite);  $M=1/2$.}\label{fig9}
\end{figure}

\begin{figure}
\centerline{\resizebox{\linewidth}{!}{\includegraphics*{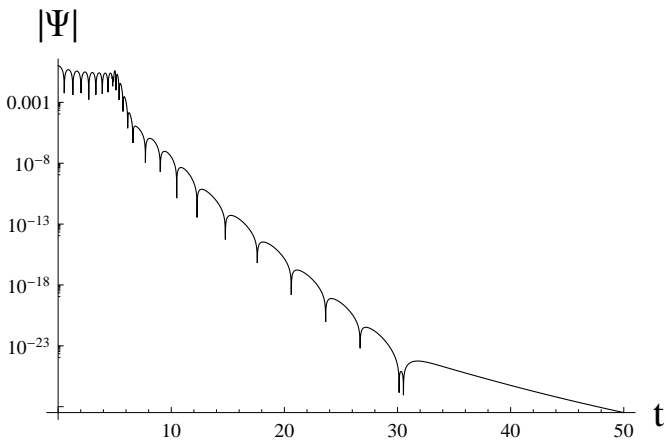}}}
\caption{Time-domain profile for the scalar type of gravitational perturbations, $\ell=10$, $\alpha =-3.99$. There are two concurrent modes, one of which dominates at the beginning of the quasinormal ringing ($\omega = 8.25 - 6.12 i$) and the other one - at the final stage ($\omega = 1.12 - 1.77 i$); $M=1/2$. The initial outburst is prolonged because the spacial point at which the time-domain profile is taken is at some distance from the peak of the effective potential.}\label{fig-add1}
\end{figure}

For negative $\alpha$ the effective potential are positive definite up to some moderately large absolute value of the coupling constant. Thus the effective potential for the scalar type of gravitational perturbations at $\ell \geq 2$ is positive definite, when
\begin{equation}
 \alpha \gtrapprox -8 M^2 \quad (scalar~ type),
\end{equation}
while for vector type of gravitational perturbations the effective potential acquires a negative gap near the event horizon at $\alpha \lesssim - 15.8 M^2$. Nevertheless, the time-domain profiles of near extremal states for sufficiently high multipoles up to $\ell =10$ show no instability. The stability of vector type of gravitational perturbations can also be shown via the S-deformations in a similar fashion with \cite{Takahashi:2010gz}. Therefore, we conclude that \textbf{the} vector type of gravitational perturbations is stable whenever
\begin{equation}
\alpha  \gtrapprox - 16 M^2 \quad (vector~ type).
\end{equation}
Examples of positive definite effective potential and potential with a negative gap near the potential barrier are given on fig. \ref{fig9}.
Thus, we conclude that the black hole is stable for $0 > \alpha \gtrapprox - 8 M^2$. For $\alpha < - 8 M^2$ the effective potential acquires the negative gap which, nevertheless, can sometimes be remedied at higher multipoles $\ell$, so that the full analysis of stability for $\alpha$ must be done via the through consideration of the quasinormal spectrum for all negative values of  $\alpha$. Indeed, as can be seen in fig. \ref{fig-add1} even the near extremal black hole with negative coupling constant  is gravitationally stable. In addition, analysis of stability which was made after the first version of our work appeared \cite{Cuyubamba:2020moe} comes to the same conclusion that there is no gravitational instability for negative values of $\alpha$ in the scalar channel.

Summarizing all the regions of instability and restoring the arbitrary $M$, we conclude that the black hole is stable once
\begin{equation}
-16 M^2<\alpha \lessapprox 0.6 M^2
\end{equation}
and unstable for values of $\alpha$ larger than $0.6 M^2$.

In  tables II and III one can see the fundamental ($\ell=2$, $n=0$) quasinormal modes of vector (axial) and scalar (polar) types of gravitational perturbations in the region which is proved to be free from instabilities. As one can see all the data obtained by the WKB and time-domain integration are in a very good agreement for small and moderate values of $\alpha$. When $\alpha$ is increasing, both the real oscillation frequency and damping rate decrease. At $\alpha < -1.5$ the discrepancy between the time-domain integration and WKB approaches slightly increases, because the second concurrent mode with nearby damping rate appears in the spectrum and the time domain profile consists from the two dominant modes (see fig. \ref{fig10}).When the negative coupling constant is close to its extremal value allowing for the black-hole solution the two modes have quite different damping rates and one of the modes dominate in the beginning of the quasinormal ringing, while the other one dominates in the end (see fig. \ref{fig-add1}).

\subsection{The correspondence between the eikonal QNMs and null geodesics}

It is worth mentioning that in the eikonal regime the quasinormal modes of test fields do not coincide with those for the gravitational perturbations. This is reflected in the broken correspondence between eikonal quasinormal modes and null geodesics. According to this correspondence (reported in \cite{Cardoso:2008bp}) the real and imaginary parts of the eikonal quasinormal mode must be multiples of the frequency and instability timescale of the circular null geodesics respectively.
Following Cardoso et. al. \cite{Cardoso:2008bp}, one can see that the principal Lyapunov exponent for null geodesics around a static,  spherically symmetric metric  is
\begin{equation}\label{GenLyap}
\lambda = \frac{1}{\sqrt{2}}\sqrt{-\frac{r_c^2}{f_c}\left(\frac{d^2}{dr_*^2}\frac{f}{r^2}\right)_{r=r_c}}.
\end{equation}
The coordinate angular velocity for the null geodesics is
\begin{equation}\label{angularvel}
\Omega_c = \frac{f_c^{1/2}}{r_c},
\end{equation}
where $r_{c}$ is the radius of the circular null geodesics,  $f_c$ is the metric function taken at $r_{c}$, satisfying the equation
\begin{equation}\label{circulareq}
 2f_c=r_cf'_c.
\end{equation}

Then, in a similar fashion with \cite{Konoplya:2017wot}, we observe that
\begin{equation}\label{extremum}
2 f_{i}(r_0)=r_0 f_{i}'(r_0),
\end{equation}
where $f_i$ is the function determined by (\ref{fv}, \ref{fs}) taken at the maximum of the effective potential $r_0$. Thus,  $f(r)$ does not coincide with $f_{i}(r)$, so that the position of the effective potential's extremum $r_{0}$ must not coincide with the location of the null circular geodesic $r_c$. The WKB formula for quasinormal modes is also different from the Einsteinian ones, as now it includes $f_{i}(r)$ instead of $f(r)$:
\begin{equation}\label{main}
\omega_{\rm QNM i}=\left(\ell + \frac{1}{2}\right) \sqrt{\frac{f_{i0}}{r_0^2}} -i\frac{(n+1/2)}{\sqrt{2}}
\sqrt{-\frac{r_0^2}{f_{i0}}\,\left (\frac{d^2}{dr_*^2}\frac{f_{i}}{r^2}\right )_{r_0}}.
\end{equation}
Thus we conclude that the correspondence between gravitational quasinormal modes in the eikonal regime and null geodesics is not fulfilled in our case, but it does take place for test fields, whenever the coupling constant $\alpha$ is small and there is yet a point to consider the background metric as the viable black-hole solution. This is a four-dimensional example of the principle formulated in \cite{Konoplya:2017wot}, but  illustrated there  for $D>4$ spacetimes: the eikonal quasinormal modes/null geodesics correspondence is guaranteed only for the good, from the WKB point of view, effective potentials, which are provided for minimally coupled test fields.

\subsection{Consistency of the obtained gravitational perturbations in various approaches to constructing the $4D$ theory}

Here we studied the linear stability of spherically symmetric black holes within the initial regularization approach \cite{Glavan:2019inb}. We hope that this approach could be applied for effective description of sufficiently large black holes, when the gravitational wavelength is of the order of the black-hole size and the latter is much larger than the Plank length, that is, the quantum effects can be safely neglected. The black hole solution found in \cite{Glavan:2019inb} is unambiguous since the Weyl part of the corresponding equations vanishes,
\begin{equation}
C^{\mu\rho\lambda\sigma}C_{\nu\rho\lambda\sigma}-\frac{1}{4}\delta^{\mu}_{\nu}C^{\tau\rho\lambda\sigma}C_{\tau\rho\lambda\sigma}=0,
\end{equation}
where $C_{\nu\rho\lambda\sigma}$ is the Weyl tensor \cite{Aoki:2020lig}.

The consistent Hamiltonian theory implies that the gravitational perturbations gain a correction to the dispersion relation in the ultraviolet regime due to counter terms, appearing in order to cancel divergences of the Weyl pieces \cite{Aoki:2020lig}. Although the linear perturbation equations for the full theory have not been derived yet, we notice that the Weyl tensor on the $(D-2)$-sphere does not appear in scalar-type and vector-type perturbation equations \cite{Kodama:2003kk}, indicating that the counter terms do not change the angular parts of the equations, which were used for our eikonal stability analysis. It is worth mentioning here that although eikonal instability of the Gauss-Bonnet black holes manifests itself first at large multipole number, the similar unstable behavior is observed for lower $\ell$ as well \cite{Konoplya:2008ix}, corresponding to the low-energy perturbations at the threshold of instability, for which the ultraviolet corrections can be neglected. Therefore, linear and higher-order perturbation analysis in the full theory could further limit the parametric region of stability of the $4D$-Gauss-Bonnet black holes.

\section{Radius of the black-hole shadow}

Theoretical analysis of shapes of the black hole shadows have been recently considered in a great number of papers (see, for example, \cite{Dokuchaev:2019jqq,Chang:2019vni,Tian:2019yhn,Pantig:2020odu,Chang:2020miq,Jusufi:2019ltj,Contreras:2019nih,Vagnozzi:2019apd} and references therein).
The radius of the photon sphere $r_{ph}$ of a spherically symmetric black hole is determined by means of the following function: (see, for example, \cite{Bisnovatyi-Kogan:2017kii,Konoplya:2019sns} and references therein)
\begin{equation}
h^2(r) \equiv \frac{r^2}{f(r)} \,,
\label{h2 definition}
\end{equation}
as the solution to the equation
\begin{equation}
\frac{d}{dr} h^2 (r)=0.
\end{equation}
Then, the radius of the black-hole shadow $R_{sh}$ as seen by a distant static observer located at $r_O$ will be
\begin{equation}
R_{sh} = \frac{h(r_{ph})r_O}{h(r_O)} = \frac{r_{ph}\sqrt{f(r_O)}}{\sqrt{f(r_{ph})}} \approx  \frac{r_{ph}}{\sqrt{f(r_{ph})}}\,,
\label{shadow def}
\end{equation}
where in the last equation we have assumed that the observer is located sufficiently far away from the black hole so that $f(r_O) \approx 1$.

One can easily see that in the units of the event horizon radius $r_{+} =1$, the radius of the shadow can be very well approximated by the following linear law:
\begin{equation}
R_{sh} \approx \frac{3 \sqrt{3}}{2} + 0.94 \alpha,
\end{equation}
where the first term is for the Schwarzschild's radius of the shadow. Thus, the radius of the shadow is always larger when the GB coupling is turned on.

\section{Discussion}
It is generally accepted that the Einstein-Gauss-Bonnet theory is nontrivial only in higher than four dimensional spacetimes. However, the re-scaling of the coupling constant prior to the dimensional reduction \cite{Glavan:2019inb} leads to the novel regularization method leading  to the effective solutions representing the Gauss-Bonnet correction in $4D$. Despite the regularization \cite{Glavan:2019inb} is possible not for every kind of metric, it was shown that the black-hole solution obtained in \cite{Glavan:2019inb} via regularization is also an exact solution of the well-defined theory which introduces the Gauss-Bonnet correction at a  cost of the broken diffeomorfism  \cite{Aoki:2020lig}, in the full agreement with the Lovelock theorem.

In the well-defined formulation of  \cite{Aoki:2020lig}, and in a subsequent paper \cite{Aoki:2020iwm} it was shown that the dispersion relation will be modified in the UV regime, which means that one could expect that our analysis of gravitational perturbation could be a reasonable approximation in the IR part of the quasinormal spectrum, which is of primary interest in astrophysics.
Our calculations of the effects for test fields (such as quasinormal modes of test fields and shadows) must be valid, because the background black-hole metric which we used  is the exact solution in the well-defined theory. The high frequency spectrum of gravitational quasinormal modes will be modified in \cite{Aoki:2020iwm} owing to the modified dispersion relations for gravitons.  However, the threshold of instability may be considerably affected in the full theory \cite{Aoki:2020iwm}, because the equation (\ref{21}) for gravitational perturbations must be a good approximation when the Gauss-Bonnet coupling is much smaller than the typical length scale, namely $|\alpha| \ll M^2$.

Here we have also studied quasinormal modes of scalar, electromagentic and gravitational perturbations of asymptotically flat black hole in the $(3+1)$-dimensional Einstein-Gauss-Bonnet black hole. We have shown that as to the change of the coupling constant $\alpha$, the damping rate is more sensitive characteristic than the real oscillation frequency. In addition, we have shown that unless the coupling constant is small enough, a dynamical eikonal instability occurs in the vector (axial) and scalar (polar) types of gravitational perturbations. This is similar to the instability observed for the higher dimensional Einstein-Gauss-Bonnet and Lovelock theories (see, for example, \cite{Dotti:2005sq,Gleiser:2005ra,Konoplya:2017lhs,Konoplya:2017zwo} and reference therein). The branch with negative $\alpha$ allows for stable black holes at much larger absolute values of the coupling constant than the branch with positive $\alpha$. In the regime of large negative $\alpha$ there appear two concurrent modes with close damping rates. The radius of the shadow is remarkably well described by the linear law.

Our paper can be extended in a number of ways. The $4D$ Einstein-Lovelock solution obtained as a result of the dimensional regularization \cite{Konoplya:2020qqh}
could be tested for its spectrum, stability and shadows in a similar fashion. The quasinormal modes and stability of an asymptotically de Sitter branch can be considered in a similar manner. The stability and gravitational quasinormal spectrum must also be studied in the full theory \cite{Aoki:2020lig,Aoki:2020iwm} taking into account the corrections to the dispersion relations owing to additional degrees of freedom in the UV regime.
In the forthcoming paper we will study the grey-body factors and Hawking radiation of the asymptotically flat 4D Einstein-Gauss-Bonnet black holes \cite{Konoplya:2020cbv}.

\acknowledgments{
The authors acknowledge A. Zhidenko for useful discussions. This work was supported by the 9-03950S GAČR grant. A. F. Z. also acknowledges the SU grant SGS/12/2019. R. K. acknowledges the support by the ``RUDN University Program 5-100''. We would like to thank the anonymous referees for useful criticism of the first version of this paper.}

\end{document}